\documentclass[11pt]{article}

\usepackage{amsmath,epsfig,amssymb,fancyhdr,graphicx,wrapfig,multirow,lscape}
\usepackage[top=1in, bottom=1in, left=1in, right=1in]{geometry}

\RequirePackage{ifthen}
\newboolean{withfigs}
\setboolean{withfigs}{true}

\ifthenelse{\boolean{withfigs}}
{
	\newcommand{\flexfig}[4]{
		\clearpage
		\begin{figure}[htb]
		\centering
		\includegraphics[#1]{#2}
		\vfill
		\caption{\label{#3} #4}
		\end{figure}
	}
}{
	\newcommand{\flexfig}[4]{
		\refstepcounter{figure}
		\label{#3}
		\paragraph*{Figure \ref{#3}.}
		#4\\
	}
}

\newboolean{useKenAlt}
\setboolean{useKenAlt}{true}
\newcommand{\textAlt}[2]{
	\ifthenelse{\boolean{useKenAlt}}{
		#2
	}{
		#1
	}
}

\newcommand{\wmark}{}
\usepackage{type1cm}
\usepackage{eso-pic}
\usepackage{color}
\makeatletter
\AddToShipoutPicture{%
	\setlength{\@tempdimb}{.5\paperwidth}%
	\setlength{\@tempdimc}{.5\paperheight}%
	\setlength{\unitlength}{1pt}%
	\put(\strip@pt\@tempdimb,\strip@pt\@tempdimc){%
		\makebox(0,0){\rotatebox{55}{\textcolor[gray]{0.85}%
		{\fontsize{4cm}{4cm}\selectfont{\wmark}}}}%
	}%
}
\makeatother

\newcounter{algo}

\newcommand{\comment}[1]{}

\newcommand{\gene}[1]{$\mathit{#1}$}

\newcommand{\var}[1]{\mathsf{Var}({#1})}
\newcommand{\sd}[1]{\mathsf{SD}({#1})}
\newcommand{\mean}[1]{\mathsf{E}({#1})}
\newcommand{\bigO}{\mathcal{O}}
\newcommand{\abs}[1]{\left|{#1}\right|}

\begin{document}


\title{Pathways of Distinction Analysis:\\ A new technique for multi-{SNP} analysis of {GWAS} data}

\author{Rosemary Braun and Kenneth Buetow\\ {\footnotesize{\textit{National Cancer Institute, NIH, Bethesda, MD.}}}}

\date{March 4, 2011}

\maketitle

\begin{abstract}
Genome-wide association studies have become increasingly common due
to advances in technology and have permitted the identification of
differences in single nucleotide polymorphism (SNP) alleles that
are associated with diseases.   However, while typical GWAS analysis
techniques treat markers individually, complex diseases (cancers,
diabetes, and Alzheimers, amongst others) are unlikely to have a
single causative gene.   There is thus a pressing need for multi-SNP
analysis methods that can reveal system-level differences in cases
and controls.

Here, we present a novel multi-SNP GWAS analysis method called
Pathways of Distinction Analysis (PoDA).  The method uses GWAS data
and known pathway-gene and gene-SNP associations to identify pathways
that permit, ideally, the distinction of cases from controls.  The technique
is based upon the hypothesis that if a pathway is related to disease
risk, cases will appear more similar to other cases than to controls
(or vice versa) for the SNPs associated with that 
pathway.  By systematically applying the method to all pathways
of potential interest, we can identify those for which the hypothesis
holds true, i.e.,  pathways containing SNPs for which the samples
exhibit greater within-class similarity than across classes.
Importantly, PoDA improves on existing single-SNP and SNP-set
enrichment analyses in that it does not require the SNPs in a pathway 
to exhibit independent main effects.  This permits PoDA to reveal pathways
in which epistatic interactions drive risk.

In this paper, we detail the PoDA method and apply it to two GWA studies:
one of breast cancer, and the other of liver cancer.  The results obtained
strongly suggest that there exist pathway-wide
genomic differences that contribute to disease susceptibility.
PoDA thus provides
an analytical tool that is complementary to existing techniques and
has the power to enrich our understanding of disease genomics at the
systems-level.
\end{abstract}

\section*{Author Summary}
We present a novel method for multi-SNP analysis of genome-wide
association studies.  The method is motivated by the intuition that
if a set of SNPs is associated with disease, cases and controls
will exhibit more within-group similarity than across-group similarity
for the SNPs in the set of interest.  Our method, Pathways of
Distinction Analysis (PoDA), uses GWAS data and known pathway-gene
and gene-SNP associations to identify pathways that permit the
distinction of cases from controls.  By systematically applying the
method to all pathways of potential interest, we can identify
pathways containing SNPs for which the cases and controls are
distinguished and infer those pathway's role in disease.  We detail
the PoDA method and describe its results in  breast and liver cancer
GWAS data, demonstrating its utility as a method for systems-level
analysis of GWAS data.

\section*{Introduction}

Genome-wide association studies (GWAS) have become a powerful and
increasingly affordable tool to study the genetic variants associated
with disease.  Modern GWAS yield information on millions of
single nucleotide polymorphism (SNPs) loci distributed across the
human genome, and have already yielded insights into the genetic
basis of complex diseases~\cite{HIRS05,MCCA08}, including diabetes,
inflammatory bowel disease, and several cancers~\cite{EAST08,CGEMS07,
CGEMS09,LOU09,THOM09}; a complete list of published GWAS can be
found at the National Cancer Institute--National Human Genome
Research Institute (NCI-NHGRI) catalog of published genome-wide
association studies~\cite{HIND09}.

Typically, the data produced in GWAS are analyzed by considering each
SNP independently, testing the alleles at each locus for association
with case status; significant association is indicative of a nearby
genetic variation which may play a role in disease susceptibility.
Genomic regions of interest may also be subject to haplotype analysis,
in which a handful of alleles transmitted together on the same
chromosome are tested for association with disease; in this case,
the loci which are jointly considered are located within a small
genomic region, often confined to the neighborhood of a single gene.

Recently, however, there has been increasing interest in multilocus,
systems-based analyses.  This interest is motivated by a variety of 
factors.  First, few loci identified in GWAS have large effect sizes
(the problem of ``missing heritability'') and it is likely that
the common--disease, common--variant hypothesis~\cite{SCHO09,MOOR10} 
does not hold in the case of complex diseases.  Second, single
marker associations identified in GWAS often fail to replicate.  This 
phenomenon has been attributed to underlying epistasis~\cite{GREE09},
and a similar problem in gene expression profiling has been mitigated
through the use of gene-set statistics.  Most importantly, it is
now well understood that because biological systems are driven by
complex biomolecular interactions, multi-gene effects will play an
important role in mapping genotypes to phenotypes; recent reviews
by Moore and coworkers describe this issue well~\cite{MOOR10,MOOR03}.
Additionally, the finding that epistasis and pleiotropy appear to be 
inherent properties of biomolecular networks~\cite{TYLE09} rather
than isolated occurences motivates the need for systems-level
understanding of human genetics. 

\comment{
However, in the case of complex diseases---including cancers, where
we will focus our attention---single causative genes are rare.
Rather, the action of several genes is required to produce the
phenotype, an effect referred to as epistasis.  Conversely, alterations
to different, distinct genes may result in the same disease
by targeting the same biological process.  As a result, even in
histologically identical tumors, only a fraction may share the same
set of mutations (see references in \cite{HANA00} for examples),
and the genomic changes underlying the disease will be missed in
the analyses.}  

The impact that biological interaction networks have on our ability
to identify genomic causes of complex disease is readily apparent.
Consider a biologically crucial mechanism with several potential
points of failure, such that an alteration to any will confer disease
risk.  Because no single alteration is predominant amongst cases,
none attain a significant association; indeed, it has long been
observed that even in histologically identical tumors, only a
fraction may share the same set of mutations (see references in
\cite{HANA00} for examples).  Additionally, in a robust system,
multiple alterations may be necessary to alter the activity of an
interaction network; here, healthy individuals may
share a subset of the deleterious alleles found in cases, and again
these loci will not be detected. This complexity, noted
by~\cite{MOOR10,MOOR03,TYLE09,HANA00} and others, has generated
considerable interest in multi-locus analysis techniques that take
advantage of known interaction information.

Several multi-SNP GWAS analysis approaches have been described
in the literature.  Thorough reviews are provided in~\cite{HOLM10,WANG10},
and we briefly describe several here.
Building on the well-established Gene Set Enrichment
Analysis~\cite{GSEA05} method initially developed for gene expression data,
two articles have proposed an extension of GSEA for SNP
data~\cite{WANG07,HOLD08}.  In these techniques, each SNP is assigned
a statistic based on a $\chi^2$ test of association with the
phenotype; a running sum is then used to assess whether large
statistics occur more frequently amongst a SNP set of interest than
could be expected by chance.  While GSEA-type approaches have proven
quite useful, their reliance on single-marker statistics means that
systematic yet subtle changes in a gene set will be missed if the
individual genes do not have a strong marginal association.  In the case
of a purely epistatic interaction between two SNPs in a set, the set
may fail to reach significance altogether.   

To address this issue, Yang and colleagues proposed SNPHarvester~\cite{YANG08},
designed to detect multi-SNP associations even when the marginal
effects are weak.  To reduce the search space of possible multi-SNP
effects, SNPHarvester~\cite{YANG08} first removes any SNPs with univarite
significance.  Using a novel searching algorithm, they identify groups
of  $l$ SNPs that show association with status in a $\chi^2$ test with $3^l-1$
degrees of freedom.  
While this approach can reveal epistatic effects that would be
missed by the GSEA-type schemes~\cite{WANG07,HOLD08}, it
has other drawbacks.  First, the combinatorial explosion of SNP groups
puts a limit on the number of SNPs $l$ that may simultaneously be 
examined.  Second, the the arbitrary
groupings of SNPs and the exclusion of SNPs with marginal effects
can make the biological interpretation of the analysis results
difficult.

The notion that cases will more closely resemble other cases than
they will controls has motivated a number of interesting distance-based
approaches for detecting epistasis.  
Multi-dimensionality reduction (MDR) has been proposed and applied
to SNP data~\cite{MOTS06, MOOR06, CORD09}.  In this technique,
sets of $l$ SNPs are exhaustively searched for combinations
that will best partition the samples by examining the $3^l$ cells
in that space (corresponding to homozygous minor, heterozygous, or
homozygous major alleles for each locus) for overrepresentation of
cases.  While this method both finds epistatic interactions without
requiring marginal effects and can be structured to incorporate
expert knowledge, it is limited by the fact the the total number of loci
to be combinatorially explored must be restricted to limit computational
cost.  To address this, an ``interleaving'' approach in which models
are constructed hierarchically has been suggested~\cite{MOOR06} to reduce
the combinatorial search space.
A recent and powerful MDR implementation~\cite{GREE10} taking advantange of the CUDA parallel
computing architecture for graphics processors has made feasible
a genome-wide analysis of pairwise SNP interactions.  Still, MDR remains
computationally challenging, such that expanding the search to other
SNP set sizes (rather than restricting to pairwise interactions) can be
impeded by combinatorial complexity if an exhaustive search is to
be performed.

In order to narrow down the combinatorial complexity of discovering
SNP sets using techniques such as MDR, feature selection may be
employed.  Of particular importance here is the distance-based approach
of the Relief family of algorithms~\cite{RELIEF, RELIEFF, TURF, SURF}.
These are designed to identify features of interest by
weighting each feature through a nearest-neighbor approach.  The weights 
are constructed in the following way: for each attribute, one selects samples
at random and asks whether the nearest neighbor (across all attributes) from
the same class and the nearest neighbor from the other class have the same
or different values from the randomly chosen sample.  Attributes for which 
in-class nearest neighbors tend to have the same value are weighted more 
strongly.   Because the distances are computed across all attributes,
Relief-type algorithms can identify SNPs that form part of an epistatic 
group and they provide a means of filtration that does not have the drawbacks
of other significance filters.

While these methods have so far been applied to finding small groups
of interacting SNPs, one may instead be interested in whether
cases and controls exhibit differential distance when considering
a large number of genes.  
A multi-SNP statistic has been proposed in the
literature~\cite{HOME08,BRAU2009,VISS2009} for determining whether
an individual of interest is on average (across a large number
of SNPs) ``closer'' to one population sample than to another.  
The method, originally proposed by Homer~\cite{HOME08}, is motivated
by the idea that a subtle but systematic variation across a large number of 
SNPs can produce a discernible difference in the closeness of an individual
to one population sample relative to another.  While this statistic
was originally designed to identify the proband as a member of one
of the population samples, it was shown in~\cite{BRAU2009} that
out-of-pool cases from a case-control breast cancer study were in
general closer (as defined by the statistic presented in~\cite{HOME08})
to in-pool cases than they were to in-pool controls, suggesting that the
combination of multiple alleles has the potential to distinguish
cases from controls.

Building on these ideas, we present a new technique that finds
pathway-based SNP-sets that differentiate cases from controls; we
call this method Pathways of Distinction Analysis (PoDA). 
%
In PoDA, SNP sets are defined based on known relationships (e.g., SNPs
in genes sharing a common pathway), and thus incorporate expert
knowledge to reduce the search space and provide biological
interpretability.  Motivated by the differential ``closeness'' of
cases and controls as discussed about and as observed in~\cite{BRAU2009},
we hypothesize that if the SNPs come from a pathway that plays a
role in disease, there will be greater in-class similarity than
across-class similarity in the genotypes for those SNPs;
i.e., a case will show greater genetic similarity to other cases
than to controls for the SNPs on a disease-related pathway, but will
be equidistant for the SNPs on a non-disease-related pathway.  Based
on this notion, PoDA seeks to identify pathways for which
differential heterogeneity is exhibited in cases and controls.
In each pathway, PoDA returns a statistic $S$ for each sample that
quantifies that sample's distance to the remaining cases relative
to its distance to the remaining controls for a given pathway's
SNPs.  PoDA then examines whether the
distributions of $S$ for the controls differ from those of the cases by
computing and testing for significance a Pathway Distinction Score $DS$
that quantifies the differences in case and control $S$ distributions.
In this manuscript, we detail the PoDA method 
and report the results of its application to two data sets.

As we will describe, PoDA improves and complements existing approaches
in a number of respects.  First, it permits the investigation of
arbitrarily large pathways, circumventing the dimensionality issues
that are encountered with MDR and SNP-Harvester.  Second, it is
able to detect pathways that contain an over-abundance of
highly-significant markers as well as pathways whose markers have
a small but consistent association that would be missed by GSEA-type
approaches.  Finally, it uses a leave-one-out technique to return
for each sample an unsupervised relative distance statistic that
can then be used to model disease risk via logistic regression.  In
addition to providing an effect size for the pathway, this allows the
odds of disease for new samples to be obtained by computing
its relative distance statistic with respect to the known samples
and applying the model. 

\section*{Methods}
Following our conjecture that SNPs associated with the genes in a
pathway involved in disease will exhibit more within-group similarity
than across-group similarity, we propose Pathways of Distinction
Analysis (PoDA), a method designed to address
the following questions:
\begin{itemize}
\item Given some set of SNPs, do we find that, on average, cases
are ``closer'' to other cases than to controls (or that controls
are ``closer'' to other controls than to cases)?

\item If we look for these distinctions systematically over all
SNP-sets of potential interest, can we use it to single out SNP-sets
which may be associated with disease?  
\end{itemize}

In PoDA, a set of SNPs are selected, and for each sample we compute
whether it is closer to the pool of remaining cases or controls
across that SNP set,
using the relative distance statistic described below.  Once this
is done for every sample, the distribution of the relative distance
statistic is compared in the cases and controls using a nonparametric
statistic, addressing the first question above.  This may be carried out
amongst all SNP sets of interest, adjusting the $p$-value for the
multiple hypotheses, to find SNP sets for which cases more strongly
resemble the population of remaining cases while controls more
strongly resemble the population of remaining controls.

We begin with a discussion of how we measure the relative distance
of an individual to the other cases vs. other controls.  A simple
but computationally intensive approach is to represent each sample
by a vector in an $l$-dimensional space, where $l$ is the number
of SNPs in the group of interest.  One can then compute, between
each sample pair, their distance in this $l$-dimensional space using
a Euclidean, Manhattan, or Hamming metric.  For each sample, we would
define its relative distance statistic as the mean of its distance to other
controls minus the mean of its distance to other cases; a sample that is
more similar to cases will exhibit a positive statistic, whereas
one that is more similar to other controls will exhibit a negative
statistic.  For the given SNP set, we would then have for each sample
a value quantifying its relative distance that was computed without
knowledge of that sample's class (i.e., using a leave-one-out scheme)
and could then be used in further tests.  By doing this for all 
pathways of interest, one derives a relative distance value for each
sample in each pathway.

This brute-force approach, while conceptually clear, has two
significant drawbacks. The first is that the distance computation
is $\bigO(l\cdot n^2)$ where $n$ is the total number of samples in the
study---a considerable undertaking, particularly if many SNP sets
are to be analyzed, and one that becomes exceedingly troublesome
in the context of permutation tests.  The second drawback is that
because we are taking the mean of the distances, a sample that is situated
squarely within a cluster of cases may have a large case-distance
value due to the dispersion of cases around it.  Both of these
issues are circumvented by instead considering the relative distance
to the \textit{centroids} of the cases and controls in the
$l$-dimensional space, a computation that can be performed in
$\bigO(l\cdot n)$ for all $n$ samples.  It is this approach that PoDA
employs, as follows:

In~\cite{BRAU2009,HOME08}, the authors consider a measure of
individual $Y$'s distance to two population samples $F$ and $G$ at
locus $i$, 
\begin{equation} 
D_{Y,i}  = \abs{y_i - f_i} - \abs{y_i - g_i}  \, .  
\label{craigD} 
\end{equation} 
where $f_i$ and $g_i$
are the minor allele frequencies (MAFs) of SNP $i$ in samples $F$
and $G$, and $y_i \in \{0, 0.5, 1\}$ is $Y$'s genotype at $i$
corresponding to homozygous major, heterozygous, and homozygous
minor alleles, respectively (i.e., the frequency of minor allele
in that individual.  The first term quantifies how different $Y$'s
MAF is from $F$'s for a given locus $i$; the second term quantifies
how different $Y$'s MAF is from $G$'s at locus $i$; and so $D_{Y,i}$
gives the distance of $Y$ relative to $F$ and $G$ at locus $i$.  
Since the minor allele frequencies $f_i$ and $g_i$ are computed by
averaging the genotypes (again, written as $\{0, 0.5, 1\}$) in samples 
$F$ and $G$ respectively, it is clear that $\abs{y_i - f_i}$ is the
distance from $Y$ to the centroid of $F$ along the coordinate $i$
(and likewise for the $g_i$ term).
It can be seen from Eq.~\ref{craigD} that positive $D_{Y,i}$ implies
that $y_i$ is closer to $g_i$  than to $f_i$, and that negative
$D_{Y,i}$ implies that $y_i$ is closer to $f_i$ than to $g_i$. 

By computing $D_{Y,i}$ at each locus $i$ and taking the standardized
mean across the $l$ loci, \cite{HOME08} obtain a distance score $S$ which
quantifies how close $Y$ is relative to $F$ and $G$ across all $l$
loci under consideration, 
\begin{equation} S_Y =
\frac{\mean{D_{Y,i}}}{\sqrt{\var{D_{Y,i}}/l}} \, , 
\label{craigT}
\end{equation} 
where $\mean{D_{Y,i}}$ denotes the mean of $D_{Y,i}$
across all loci $i$.  That is, $S$ provides a means to quantify whether
$Y$'s MAFs are closer to $G$'s MAFs or $F$'s MAFs on average for
the loci under consideration.  It is instructive to consider the 
geometrical interpretation of Eq.~\ref{craigT}. Is clear upon inspection that the numerator in Eq.~\ref{craigT} is equal, up to a factor of $l$, to the
difference in Manhattan distances between $Y$ and the (nonstandardized)
$G$ centroid and $Y$ and the (nonstandardized) $F$ centroid; in this sense,
Eq.~\ref{craigT} resembles a nearest-centroid classifier.  However, the
denominator scales the relative distances by their variance across the $l$
SNPs; that is, a sample $Y$ who is consistently closer to $G$ than to $F$
for each of the $l$ SNPs will obtain a higher $S$ than an
individual who is variously closer to either across the $l$
SNPs under consideration. 

By assigning the (non-$Y$) controls as $F$ and the (non-$Y$) cases
as $G$, we can compute a statistic $S_Y$ quantifying $Y$'s distance
to other cases relative to $Y$'s distance to other controls.  If
we then apply this systematically to all individuals in the study
population (removing that individual, computing the MAF's $f_i$ and
$g_i$ for the remaining individuals who comprise $F$ and $G$, and
then computing $S_Y$ in a leave-one-out manner), we can obtain
distributions of $S_Y$ statistics in cases and controls that may
be compared.  Here, the null hypothesis is that case and control
$S_Y$ distributions do not differ, with the alternative hypothesis
that the cases have higher $S$ values than do controls, i.e.,  that
they are closer (via Eqs.~\ref{craigD}-\ref{craigT}) to other cases
than are controls.

We can use $S$ in the following manner to answer the questions
posed above by applying it in a leave-one-out manner in each pathway:
\begin{enumerate}
\item For a given pathway $P$, select the $l_P$ SNPs associated with that pathway;
\item For every sample $Y$, remove $Y$ from the case or control group as needed, and compute $S_{Y,P}$ with respect to the remaining cases and controls using the SNPs chosen in step 1.
\item Quantify the differences in distribution of $S_{Y,P}$'s for the case samples versus that of the controls and test for significance.
\end{enumerate}
By systematically carrying out the above steps on all pathways of
interest, we can identify pathways for which there appears to be
differential homogeneity in cases and controls, indicating that the
pathway may play a disease-related role.  The details of the 
algorithm are explained below, and summarized in Table~\ref{alg1}.

In~\cite{BRAU2009}, we examined Eqs.~\ref{craigD}-\ref{craigT} and
found that the magnitude of $S$ is influenced both by the MAF
differences $f_i-g_i$ (that is, how distant the centroids of $F$
and $G$ are) and by correlations between the SNPs under consideration
(due to the penalization for variance in $D_i$ provided by the
denominator of Eq.~\ref{craigT}.  These properties are extremely
well-suited to the application we propose: pathways with few
highly-significant SNPs will yield large $S$ differences (due to
the influence of $f_i-g_i$), as will pathways with SNPs that are
highly correlated yet have subtle individual MAF differences,
corresponding to the concerted action of multiple SNPs.

At the same time, however, we wish to ensure that the pathways
we select as having differential $S$ are not being influenced
large LD blocks covered by the SNPs in the genes on the pathway.
That is, we wish to ensure that the SNP correlations which drive $S$ are
reflective of epistatic effects between different genes rather than
recombination events within a gene.  To this end, we select a 
single SNP to represent each gene, based on the desire to detect
multi-\textit{gene} rather than multi-SNP effects.

In practice, SNPs are selected as follows: for each pathway represented
in the Pathway Interaction Database~\cite{PID} (PID, \texttt{http://www.pid.nci.gov},
containing annotations from BioCarta, Reactome, and the
NCI/Nature database~\cite{PID}) and KEGG~\cite{KEGG}, we select the associated genes.
Using dbSNP~\cite{dbSNP}, we retrieve the SNPs associated with the
pathway genes that are present in the data, excluding those with
$>20\%$ missing data or with minor allele frequency $<0.05$ in
either case of control group.  We necessarily exclude pathways for which only
one gene is probed by the remaining SNPs.  Because we are interested
in $S$ values that are driven by correlations 
\textit{across} genes (and not in individual genes covered by many SNPs 
with high LD), we select for each gene its most significant SNP in a
univariate test of association (Fisher's exact test).  It should be
noted here that while the SNP chosen for each gene is the most
significant of that gene's SNPs, it is \textit{not} necessarily
significantly associated with disease.  Our goal here is not to 
filter based on SNP significance, but rather to select, for each
gene, a single marker that is as informative as possible.

Having selected the SNPs of interest, we compute $D_{Y,i}$ at each
locus for every sample by selectively removing it and comparing it
to the remaining cases and controls, as described above.  For each
pathway $P$, we compute $S_{Y,P}$ for $l_P$ the SNPs $i$ that comprise it,
yielding a distribution of $S_{Y,P}$ for cases and another distribution
for controls.  The difference in the location of the case and control 
$S_{Y,P}$ distributions is then quantified nonparametrically by computing
the Wilcoxon rank sum statistic, defined as
\begin{equation}
W_P = \sum_{Y\in\mathrm{case}} R_{Y,P} - \frac{n_\mathrm{case}(n_\mathrm{case}+1)}{2} \,
\label{wilc}
\end{equation}
where $R_{Y,P}$ is the rank of $S_{Y,P}$ amongst all samples $Y$ for a given
pathway $P$.  Eq.~\ref{wilc}
thus quantifies, non-parametrically, the degree to which cases are ``closer''
to other cases and controls ``closer'' to other controls across a set
of SNPs for all individuals in the GWAS.

\comment{
To illustrate the above, we consider $1000$ cases and $1000$
controls and a pathway with $l=12$ SNPs.  Alleles were simulated
as binomial samples from source populations with slightly different
MAFs for cases and controls ($\sim0.22$ vs $\sim0.20$), such that
the cases were slightly more likely to be homozygous minor than were
the controls (and hence more similarity within-groups than across-groups).
The difference in the MAFs of the resulting samples
was not substantial enough to yield $p\leq 0.05$ under a Fisher's
exact test for any of the 12 individual SNPs, yet---as can been seen in
Figure~\ref{exres}(a)---the accumulated effects yield significantly
different distributions of $S$ for the two groups. By contrast, a
null data set in which the cases and controls were drawn from the
same source population yielded no significant shift in the case $S$'s
compared to controls, Figure~\ref{exres}(b).  
}

To illustrate the above, we consider a simulated GWAS of $250$ cases
and $250$ controls and $50$ SNPs, shown in Figure~\ref{exres}, and
ask whether we are able to detect a 12-SNP pathway in which a subset
of SNPs appear to have an epistatic interaction.  Alleles were
simulated as binomial samples from a source population with MAFs
ranging from $0.1$ to $0.4$ across the $50$ SNPs, and case labels
were assigned such that a combintion of homozygous minor alleles
at SNPs 1 and 2 or 3 (i.e., $(y_1=1) \land ((y_2=1) \lor (y_2=1))$)
conferred a three-fold relative risk, mimicking an epistatic
interaction between SNPs 1 and 2 and SNPs 1 and 3 (Figure~\ref{exres}(a)).
Alone, none of the $50$ SNPs showed any association with case status,
nor was any SNP significantly out of HWE in either cases or controls.
However, the ``shared pattern'' in SNPs 1--3 is such that a 12 SNP
pathway comprising SNPs 1--12 yields greater $S$ in cases than in
controls as can been seen in Figure~\ref{exres}(b), while a random
12 SNP pathway selected from the 50 SNPs (that includes SNP 3, but
neither SNP 1 or 2) shows no difference in $S$ values as seen in
Figure~\ref{exres}(c).

%
While the Wilcoxon statistic $W$ is normal in the large-sample limit
and can be directly compared to a Gaussian, to truly evaluate the
significance of $W_P$ for a given pathway $P$, we must
address two sources of bias: the number of SNPs per gene, and the 
size of the pathway.  To address these issues, we introduce a normalized
Pathway Distinction Score $DS_P$ that we test for significance using
a resampling procedure.

First, we expect that because we have selected for each gene the
single most informative SNP, we are pre-disposed to seeing higher $W_P$
for pathways that contain large genes. 
Because large genes will be more likely to contain highly-significant
SNPs by chance, the concern has been raised that \cite{KRAF09,WANG07}
selecting the single most significant SNP as a proxy for the
gene (as is done here) will lead to a bias toward pathways that
contain an abundance of large genes.  To account for this, we
follow the approach in~\cite{WANG07} and normalize the score via
a  permutation-based procedure.  First, we permute
the phenotype labels and in each permutation recalculate $W_P$ as described 
above, but using the permuted case and control labels. The permuted labels 
are used both to select the most
informative SNP per gene and to compute $f_i$, $g_i$, and $W_P$ in
Eqns.~\ref{craigD}--\ref{wilc}).  This yields a distribution of $W^*_P$
under the null hypothesis that the magnitude of $W$ is independent of the
true case/control classifications.
We then normalize the true $W_P$ by the distribution from the permutation
procedure, yielding a Distinction Score $DS_P$ for pathway $P$
that effectively adjusts for different sizes of genes and preserves 
correlations of SNPs in the same gene:
\begin{equation}
DS_P = \frac{W_P - \mean{W^*_P}}{\sd{W^*_P}} \, ,
\label{poda}
\end{equation} 
where $W^*_P$ are the set of $W_P$ obtained for pathway $P$ across the
permutations.  (In practice, 100 permutations are used.) 
Because the permuted labels are used in the SNP selection, this 
normalization adjusts for the bias introduced by the fact 
that large genes have more opportunity to contain significant SNPs
by chance.  The Pathway Distinction Score $DS_P$ thus provides a
model-free, gene-size adjusted metric for quantifying the degree
to which cases are ``closer'' to other cases (higher $S_P$) than
controls.

To test whether $DS_P$ is significant, we note that larger
pathways may yield high $DS_P$ values simply due to the fact that they
sample the case anc control differences more thoroughly.  Indeed, the question of
significance that we wish to address is not simply whether a pathway
permits the distinction of cases and controls, but \textit{whether
it does so better than a random collection of as many SNPs}, wherein
the SNPs are still selected to be the most informative by gene.
To account for the fact that the pathways are of differing sizes,
significance of the Distinction Score for a given pathway is
assessed through resampling by choosing, at random, the same number
of SNPs that are present in that pathway ($l_P$) from the total set of
most-informative-SNP-per-gene and recomputing $DS_P$ for the random
pathway.  The $p$ value is obtained by
counting the fraction of random $l_P$-SNP sets which give a larger
$DS_P$ than the true pathway SNPs in $10^4$ resamplings.
In this way, we are able to detect pathways
that yield large differences of case and control $S$ distributions
due to their particular SNPs, rather than simply being the result
of choosing many SNPs.  The $p$ value obtained addresses the question
of whether the pathway under consideration permits greater separation
of cases and controls than would a random collection of
most-informative-SNP-per-gene, i.e., whether there exists a more
extreme aggregated effect in that pathway than expected by chance.

\section*{Results}
We applied PoDA to 2287 genotypes obtained from the
Cancer Genomic Markers of Susceptibility (CGEMS) breast cancer
study.  The samples were sourced as described in~\cite{CGEMS07}.
Briefly, the samples comprised 1145 breast cancer cases and a
comparable number (1142) of matched controls from the participants
of the Nurses Health Study.  All the participants were American
women of European descent.  The samples were genotyped against the
Illumina 550K arrays, which assays over 550,000 SNPs across the
genome.

We also applied  it to a smaller liver cancer GWAS~\cite{CLIF09}
comprising 386 hepatocellular carcinoma (HCC) patients and 587
healthy controls from a Korean population.  Samples were genotyped
against Affymetrix SNP6.0 arrays, which provides SNP information
at approximately one million loci.

\subsubsection*{Breast cancer GWAS results}

We begin by applying PoDA to the CGEMS breast cancer
GWAS data.  Having observed (Figure~\ref{exres}) that PoDA performs
as expected for the simulated data, we first turn our attention to
a simple test in which we select a SNP set comprising the four SNPs
in intron 2 of \gene{FGFR2} that were reported to show significant
association with case status in \cite{CGEMS07} (rs11200014, rs2981579,
rs1219648, rs2420946). We expect to see a strong difference in the
test case and test control distributions, and indeed we do: the
cases more frequently have positive $S$ than do controls in
Fig.~\ref{multisnp-4fgfr2}.  (The discrete peaks in the distribution
are a result of the fact that with four SNPs there exist fewer
available values of $S$.)  Using a nonparametric Wilcoxon rank sum
test with the alternative hypothesis that cases have greater $S$
than controls, $p=1.016\cdot10^{-6}$ is obtained, confirming our
intuition.

We next applied PoDA systematically to the pathways represented in
PID~\cite{PID} using CGEMS data.  Associations between genes and
SNPs were made using dbSNP build 129~\cite{dbSNP}.  1081 pathways
were non-trivially covered in the data set; 69453 SNPs in the data
could be associated with at least one of the pathways.  Because
these 69453 SNPs were associated with 4446 unique genes, 4446 were
kept in the analysis (the most significant SNP for each gene of
interest).  The 1081 pathways ranged from 2 to 229 genes, with a
mean of 19.  $S_{Y,P}$ was computed in each pathway $P$ for each
of the 2287 samples $Y$ via Eq.~\ref{craigT}, and the distinction 
score $DS_P$ (Eq.~\ref{poda}) 
quantifying differential $S$ distributions
in cases and controls was computed for each pathway.  Significance
was assessed as described above, by resampling ``dummy'' pathways of the same
length and computing the fraction of  greater $DS_P$ scores.

Because PoDA provides for each sample a measure $S$ (Eq.~\ref{craigT}
of that sample's relative distance from the remaining ones that is
obtained without regard to that sample's true class membership, we
can use the $S$ value as a metric by which to predict the odds of disease.
Here, we construct a logistic regression model of case status as a
function of $S$ to obtain the odds ratio.  $p$-values were adjusted
for the  multiplicity of pathways using FDR adjustment~\cite{FDR,
BENJ01}.  

Pathways with significant $DS_P$ and odds ratios are
reported in Table~\ref{brtab} and plots of $S$ for four of them are
illustrated in Figure~\ref{brall}.  Although the cases and controls are
not crisply separable, a unit increase in $S$ over its range from approximately
-3 to 3 yields between a 1.5 and 2.0-fold increase in odds.  Importantly,
given known minor allele frequencies for cases and controls for this set of SNPs,
we can model the increase in odds  for an unknown individual based on her 
``closeness'' to other cases.

In order to ensure that the pathways listed were not interrogating the
same set of genes, we carried out two checks.  First, we computed the SNP
overlap between all pairs of significant pathways, sequentially removing 
pathways that shared in excess of 60\% of their genes with another pathway.
Because this is done using a greedy algorithm that depends on the order of
the pathways input, the culling algorithm was run with different starting orders,
and the most frequent output was kept.  No pathway remaining in Table~\ref{brtab}
shares more than 60\% of its SNPs with another pathway.  (An un-culled list may
be found in Supplementary Table~\ref{brsup}.)  Second, we computed the
correlation of $S$ values between each pair of pathways to assess whether any pathway's
$S$ statistic was reflecting the same genetic variation as another (i.e., whether
samples that had high $S$ values for one pathway consistently did so in another).
The maximum correlation of $S$ values observed between any two pathways in
Table~\ref{brtab} was 0.58, suggesting that a different subset of
samples is affected in each pathway.

\comment{
Many of the pathways listed in Table~\ref{brtab} fulfill biological
functions that are well known to be cancer-associated, playing a
strong role in cell proliferation and migration, processes which
are perturbed in malignancies.  Focal adhesion---the
most significantly associated pathway---has a well-established 
connection with cancer and is already being targeted
by novel cancer therapeutic drugs~\cite{WEIN93,MCLE05,CHAT07}.  
Purine metabolism has been observed to be altered in cancer
cells~\cite{WEBE77,WEBE83}, and the connection of folate (a necessary
element of purine sythesis) to breast cancer is known~\cite{LEWI06,ALLE87}.
In particular, the majority of significant pathways are directly
related to cell migration (including focal adhesion, regulation of
actin cytoskeleton, gap junction, and axon guidance) and cellular
signalling (including calcium signaling pathway, PKC-catalyzed
phosphorylation of myosin phosphatase, attenuation of GPCR signaling,
and activation of PKC through GPCRs), all of which have been implicated in
a variety of cancers, while eicosanoids and unsaturated fatty acid
metabolism have been associated with breast cancer
specifically~\cite{ROSE90}.  In general, the findings in Table~\ref{brtab}
suggest that there exist germline genetic differences in these
mechanisms that confer a predisposition to disease.
}

Many of the pathways listed in Table~\ref{brtab} fulfill biological
functions that are well known to be cancer-associated, playing a
strong role in cell proliferation and migration, processes which
are perturbed in malignancies.  Purine metabolism---the most
significantly associated pathway---has been observed to be altered
in cancer cells~\cite{WEBE77,WEBE83}, and the majority of the other
significant pathways are directly related to cell migration (e.g.,
ErbB signaling and gap junction pathways) and cellular signalling
(e.g., calcium signaling, PKC-catalyzed phosphorylation of myosin
phosphatase, attenuation of GPCR signaling, and activation of PKC
through GPCRs) processes that have been implicated in a variety of
cancers. In addition, eicosanoids and unsaturated fatty acid
metabolism have been associated with breast cancer
specifically~\cite{ROSE90}. In general, the findings in Table~\ref{brtab}
suggest that there exist germline genetic differences in these
mechanisms that confer a predisposition to disease.

Interestingly, the GnRH (gonadotropin releasing hormone) signaling
pathway appears to be significant.  GnRH has been linked with
HR-positive breast cancer and the use of GnRH analogues in breast
cancer treatment has already been proposed~\cite{EIDN87,MANN86}.
However, a recent large sequencing study found no association of
GnRH1 or GnRH receptor gene polymorphisms with breast cancer
risk~\cite{CANZ09}, contrary to the author's hypothesis that common,
functional polymorphisms of GnRH1 and GnRHR could influence breast
cancer risk by modifying hormone production.  In contrast to their null 
findings, our result suggests that there are system-wide variations in GnRH
signalling that contribute to risk that are not evident when
considering the GnRH1 and GnRHR SNPs independently.

Of the 1081 pathways considered, four---FGF signaling,
MAPK signaling, regulation of actin cytoskeleton, and prostate
cancer---contained \gene{FGFR2}, the gene found to be
significantly associated in the initial CGEMS analysis~\cite{CGEMS07}.
However, only one---prostate cancer---was significant 
in comparison to randomly generated pathways of the same length.
It may reasonably be asked, then, whether the high 
significance of the prostate cancer pathway in Table~\ref{brtab}
is a result of \gene{FGFR2}.  To address this, we eliminated the
\gene{FGFR2} SNP from the prostate cancer pathway; the resampling-based
test remained significant ($p(DS_P)=0.044, OR=0.3, q(OR)=$ 8.2e-09),
suggesting that the association of the prostate cancer pathway is
not driven solely by differences in \gene{FGFR2}.

\comment{
Finally, we provide in Table~\ref{brtab} the $p$ values for the
Kolmogorov Smirnov test interrogating whether a pathway oversamples
large genes, as described in Methods.  Although some of the pathways
do oversample large genes, many do not, and we have chosen to
continue to count as PoDA-significant even those pathways that
over-sample large genes on the basis of the assumption that a pathway
with large genes will be more susceptible to genomic hits and will
remain biologically relevant.
}

\subsubsection*{Liver cancer GWAS results}

We carried out the same procedure in using data from the liver
cancer GWAS described above.  Here,
1049 pathways were non-trivially covered in the data set;
53079 SNPs in the data could be associated with at least one of the pathways.
Because these 53079 SNPs were associated with 3718 unique genes, 3718 were
kept in the analysis (the most significant SNP for each gene of interest).
The 1081 pathways ranged from 2 to 193 genes, with a mean of 16.  As above,
$DS_P$ scores for differential $S$ distributions in cases and controls
were computed for each pathway, resampled $p$ values obtained for each pathway 
size, odds ratios for $S$ were obtained, and the multiple hypotheses were
corrected using FDR adjustment~\cite{FDR, BENJ01}. 
Significant pathways are listed in Table~\ref{livtab}, and
plots of the top three pathways are given in Figures~\ref{livall}a-d.
As in the breast cancer data above, we removed pathways which had over
60\% their SNPs covered by another pathway (a complete list, with overlapping
pathways, is give in Supplementary Table~\ref{livsup}) and examined the correlation
in $S$ for all remaining pathways (maximum $\rho=0.42$).

The results here are interesting. First, we observe that a couple
pathways are significant in both the CGEMS breast and liver GWAS
with similar effect sizes, namely ErbB signaling and biosynthesis
of unsaturated fatty acids. ErbB has a well--established association
with cancer; unsaturated fatty acid biosynthesis may link diet to
cancer risk, and its appearance may suggest a gene-environment
interaction. The commonality of these known cancer-associated
pathways across the two studies suggest that there may exist genetic
patterns that confer carcinogenesis risk irrespective of the disease 
site. Along with those shared in the breast cancer data, many of
the other significant pathways in the liver cancer data well known
to be tumorassociated, including cell adhesion molecules, Wnt
signaling, c-Kit receptor, and angiogenesis pathways, further
supporting the notion that germline genetic differences in these
mechanisms contribute to cancer risk. The appearance of many neuronal
pathways is also supported by our understanding of carcinogenesis:
thes contain well-known signal transduction molecules including Ras
and PKA that may both be driving their conferring increased cancer
risk and driving the significance of the pathway~\cite{NAKA01}.

\comment{
Additionally, four of the 20 significant
liver cancer pathways are immune-related, namely, B cell receptor signaling,
antigen processing and presentation, complement and coagulation
cascades, and ``activation of Csk by cAMP-dependent protein kinase
inhibits signaling through the T cell receptor.''}

Additionally, six of the 25 significant liver cancer pathways are
immune-- and inflammation--related, namely, antigen processing and
presentation (two, with $<$60\% overlap), classical complement
pathway, corticosteroids, IL12 signaling mediated by STAT4, and
NO2-dependent IL-12 pathway in NK cells.
This is a particularly interesting finding in light
of the fact that the original analysis of the liver data~\cite{CLIF09}
suggested that altered T-cell activation plays a direct role in the
onset of liver cancer.  The involvement of the immune system in
liver cancer development has been established in clinical studies
and research involving model organisms.  Increased activity of
helper T-cells, which promote inflammation, is associated with
hepatocellular carcinogenesis~\cite{TSVE09} while activation and
proliferation of cytotoxic T-lymphocytes is suppressed in liver
cancers~\cite{ORMA05,UNIT05}.  The inflammatory immune response,
mediated by interleukins, has also been closely connected to liver
cancers in mice~\cite{NAUG07} and humans~\cite{BUDH05,BUDH06,BUDH06a}.
These findings, coupled with the observation of several significant
immune-related pathways in our data, are suggestive of
germline polymorphisms in immune response that lead to hepatocellular
carcinoma risk.

\comment{
Another notable pathway in the list is peroxisome proliferator
receptor \gene{PPAR\alpha} signaling. PPARs are nuclear receptor
transcription factors that regulate lipid and glucose metabolism.
Although the role of \gene{PPAR\alpha} in hepatocarcinogenesis in rodents
has long been established and \gene{PPAR\alpha} polymorphisms have been
linked to metabolic syndrome in humans, the connection between \gene{PPAR\alpha}
and HCC in humans remains controversial~\cite{JIAN06, PETE05}.  The 
appearance of \gene{PPAR\alpha} signaling and glycerolipid metabolism
in the list of significant pathways suggests that there may exist genetic
variations in lipid metabolism that contribute to liver cancer risk.
}

\subsubsection*{Combining pathways}

In both the breast and liver cancer results, we see observe that
even though significant pathways yield between a 1.5 and 2.0-fold
increase in odds for each unit increase in $S$ (over its typical
range of approximately $-3$ to $3$), the cases and controls are not
crisply separable based on $S$ values.
These findings suggest that it may be possible to
combine pathways to yield a model that is more predictive than a
single pathway alone.  However, the $S$ values must not simply be
put into the regression model because the overlap in pathways will
result in some SNPs being double-counted.  Rather, we combine
pathways by taking the union of their SNPs, and recomputing the
statistics.  Doing this sequentially for the top pathways in the
order as listed in Tables~\ref{brtab} and \ref{livtab}  yields the
values given in Tables~\ref{brtop} and \ref{livtop}, respectively.
Considerably higher ORs are obtained when combining the significant
pathways.  An illustration of the case and control distributions
when using a ``superpathway'' comprised of the top three pathways
in the breast and liver data, respectively, is given in Figure~\ref{top}.
These findings support the notion that the genomic variation contributing
to risk is spread over several mechanisms, rather than being concentrated
in a single gene.

\section*{Discussion}

We have introduced the Pathways of Distinction analysis method
(PoDA) for identifying pathways which can be used to distinguish
between phenotype groups.  PoDA identifies sets of SNPs in GWAS
studies for which cases and controls exhibit differential ``closeness''
to other cases and controls; that is, it permits one to infer whether
cases are more similar to other cases than are controls across a
given set of SNPs.  Because PoDA is designed to detect the joint
effects of multiple SNPs, it presents an approach to GWAS analysis
that augments single-SNP or single-gene tests.

We applied PoDA to two GWAS data sets, with highly
promising results.  In the breast cancer data, we found a number
of pathways which are known to play a role in cancers generally
and breast cancer specifically, suggesting that differences
in these mechanisms which confer disease risk may exist at the
germline DNA level.  In
the liver cancer data, we found an extreme abundance of immune-related
pathways, further corroborating the known link between inflammation
and hepatocellular carcinoma, and bolstering  the observation in~\cite{CLIF09}
that germ-line differences in immune function may play a role in
liver carcinogenesis.

PoDA may be used as a complement to other multi-SNP
analysis techniques~\cite{WANG07,HOLD08,YANG08,MOTS06}.  Unlike
gene-set enrichment type approaches~\cite{GSEA05,WANG07,HOLD08},
which search for an overabundance of significant markers in a gene
set of interest, PoDA finds both sets containing
highly significant markers as well as sets that have a subtle but
consistent pattern across all the markers in the set.  This permits
the detection of pathways in which the joint action of several alterations 
produce a phenotype and those for which any of several
possible alterations, none of them the dominant one, confer
predisposition to disease.  Indeed, many of the pathways indicated
in our analysis of the breast cancer data (Table~\ref{brtab}) were not
detected using SNP-set enrichment~\cite{GSEA05,WANG07,HOLD08} (data
not shown), including the highly significant purine metabolism and
GnRH signaling pathways, both of which are biologically relevant
(purine metabolism has been implicated in cancers generally due to
its role in DNA and RNA synthesis~\cite{WEBE83}, and GnRH has been
shown to be clinically important in breast and gynecological
cancers~\cite{EMON2003}).  These pathways, along with others that
were indicated using PoDA but not enrichment analysis
(data not shown), have a statistically significant difference
in case and control $S$ distributions and remain significant in
comparison with randomly-generated pathways of the same length.

Because PoDA effectively measures the closeness of
each individual to remaining cases and controls, it bears a conceptual
relationship to nearest-neighbor and nearest-centroid classifiers~\cite{kNN,SAM},
as well as to the distance-based feature selection algorithms like
Relief-F and its derivatives~\cite{RELIEF, RELIEFF, TURF, SURF}.
However, 
it must be remembered that the goal of PoDA is to indicate
\textit{mechanisms} that may be deleteriously hit at the genomic
level even when those hits are heterogeneous, whereas the goal of
nearest-centroid classifiers and Relief-F--type feature selection is
to derive a minimal set of markers that best classify cases and
controls (and thus are the most homogeneously hit).  These approaches
are complementary, and one can easily envision an application in
which (e.g.) Relief-F is run \textit{within} pathways that are
highly significant in the PoDA analysis in order to single out the
SNPs driving the effect.  In fact, this approach may 
improve ReliefF's ability to find those genes, since the nearest
neighbors from  which the Relief SNP weights are calculated would
be the nearest-neighbors for that specific pathway, thus discounting
heterogeneity introduced by mechanistically unrelated genes.
For instance, in the provided example (Fig~\ref{exres}), ReliefF
fails to identify the significance of SNPs 1--3 when run using
the complete 50-SNP data, but places at least two of SNPs 1, 2 or 3
in the top third of selected features when restricted to SNPs 1--12.
\comment{clustering.  
However, several features distinguish
our technique from clustering algorithms.  First, PoDA is not
entirely unsupervised; that is, in Step 2 of PoDA as presented
in Table~\ref{alg1}, class-conditional MAFs are computed, and the
comparison of $S$ distributions likewise takes place in a
class-conditional fashion.  More importantly, PoDA provides
a quantification of how much more similar cases are to other cases
than controls are, yielding a $p$ value for statistical inference
regarding whether cases are more case-similar than controls across
a chosen set of SNPs.}

While PoDA has many benefits, it should be noted that when 
epistasis drives a phenotype with \textit{no} differences in the minor
allele frequencies for the epistatically-interacting genes (as opposed to a
slight yet consistent one shown in the example),
PoDA as computed
via Eqs.~\ref{craigD},\ref{craigT} will miss the pathway.  
Geometrically,
such a situation would mean that the case and control groups have the
same centroids while having a different distribution of samples about
those centroids.  A famous example of this is provided through the
non-linearly separable XOR (exclusive or): 
consider two epistatic loci $(X,Y)$ such
that all controls have genotypes in the set $\{(0,0),(1,1)\}$ and all
cases have genotypes in the set $\{(0,1),(1,0)\}$ (i.e., that a genotype
of 1 at either locus can be compensated by a genotype of 1 at the other,
but having just one alone---1 at exclusively $X$ or $Y$---is deleterious).
If the loci $X$ and $Y$ each have the same MAF in cases and controls, it is
plain to see that the centroids will be in the same location for
both groups, and Eq.~\ref{craigD} will yield zero for both cases and controls.  
If instead of using Eq.~\ref{craigD}, we compute pairwise
sample-sample distances, we can circumvent this limitation and find such 
epistatic situations (it is this 
pairwise approach that permits Relief-F to also uncover nonlinearly
interacting SNPs).  While we provide the facility for this in the PoDA
package, the cost of carrying out the pairwise computation is a considerable 
increase in computational complexity.

A number of potential avenues exist to extend the application of
PoDA further.  One possible application is in improving the
reproducibility of GWAS results.  We note that several of the
pathways identified in the breast cancer GWAS data were also
implicated in the liver cancer data, which suggests that there may
be common features which distinguish individuals to cancer generally.
Because different GWA studies---even those of the same phenotypes---often
yield different results at the SNP level, it may be possible to
find common alterations at the pathway level across disparate 
GWAS using PoDA.

Extending PoDA further, the $DS_P$ scores obtained for
each pathway may be examined for over-representation of extreme
values in pathways that comprise a particular biological subsystem---one
may think of this as a ``pathway-set'' enrichment analysis (which
would be conducted using the a running-sum statistic analogous
GSEA~\cite{GSEA05}), and could use it to answer whether (e.g.)
immune-related pathways are hit in liver cancer more often than
expected by chance.  Alternatively, boosting~\cite{BUHL07,MEYE03}
could be used to find sets of pathways which are more predictive
of case status than individual pathways.  Either of these approaches
would yield a richer, systems-wide view of the connection between
genotype and phenotype.  Finally, because PID contains topological
information regarding pathway connectivity, one may consider
sub-networks of pathways, permitting one to find potential
chemopreventive and therapeutic targets.  Alternatively, 
Relief-F can be used, as mentioned above, in a pathway--specific
manner to yield the subset of SNPs that drive the distinction
of cases and controls in high-$DS_P$ pathways.

PoDA provides an advantage over existing GWAS analysis methods.
Because it does not rely on the significance of individual markers,
it has the power to aid in identifying the genomic causes of
complex diseases that would not be detected in single-gene tests or
enrichment analyses.  The size of the SNP set is not limited in
PoDA, and since PoDA leverages known biological relationships
to find multi-SNP effects, the results are readily interpretable. 
PoDA may thus be used to augment existing analysis techniques and
provide a richer, systems-level understanding of genomics.


\section*{Availability}
An R package to carry out PoDA is available upon request from the authors (to be deposited in the Bioconductor in the near future).

\section*{Acknowledgments}
This research was supported by the Intramural Research Program of the
National Cancer Institute, National Institutes of Health, Bethesda, MD.
RB was supported by the Cancer Prevention Fellowship Program, National
Cancer Institute, National Institutes of Health, Bethesda, MD.  The
authors would like to that Dr Carl Schaefer (NCI) for helpful discussion
and assistance with PID.


\bibliographystyle{plos}
\bibliography{/h1/braunr/Papers/tex-repo/journals_short,/h1/braunr/Papers/tex-repo/rb-nci,/h1/braunr/Papers/tex-repo/k22}

\newcommand{\multisnpcap}{
{Toy example of differential multi-SNP homogeneity in cases and controls displayed as a heat map.}
{Samples occupy columns, with cases on the right; loci occupy rows.  Homozygous minor alleles are shown in red, heterozygous in grey, and homozygous major in white.  While all loci are in HWE and have a minor allele frequency of 0.45 in both groups, a clear pattern exists amongst the first twelve SNPs in cases (see Introduction) that would not be discerned by looking at the loci individually.}
}

\newcommand{\exrescap}{
{PoDA applied to simulated data.} 
{Alleles at 50 loci for 250 cases and 250 controls were simulated such
that each SNP was in HWE and not associated with case status,
but homozygous minor (red) at both loci 1 and 2 or 1 and 3 yielded
a three-fold relative risk (a). A 12-SNP pathway comprising SNPs
1--12 shows differential $S$ distributions (b); a random 12-SNP
pathway does not (c). Boxplots are overlayed on the scatterplots
of $S$ for clarity.}
}

\newcommand{\fgfrcap}{
{PoDA applied to four highly-significant SNPs.}
{Shown is the distribution of $S$ values in CGEMS cases (red) and
controls (black) for a SNP-set comprised of four highly-significant
SNPs located in the \gene{FGFR2} gene~\cite{CGEMS07}.  
As expected, there is a substantial difference in case and control
$S$ values, with the cases having higher $S$ (i.e., closer to other
cases) than controls.  The discreteness of the distributions are
due to the fact that with four SNPs, a finite number of $S$ values
are possible.}
}

\newcommand{\brallcap}{
{Four significant pathways in breast cancer data.}
{Scatter plots
of $S_{Y,P}$ for each pathway are overlayed with boxplots are given in
the left panel; higher values of $S$ indicate that the sample is
closer to other cases than it is to other controls.  Distributions
of $S$ for cases (red) and controls (black) are given to the
right.  A significant shift toward higher $S$ values is seen in the
cases.  Odds ratios and FDR-adjusted OR $p$ values
are given.}
}

\newcommand{\livallcap}{
{Four significant pathways in liver cancer data.}
{Scatter plots
of $S_{Y,P}$ for each pathway are overlayed with boxplots are given in
the left panel; higher values of $S$ indicate that the sample is
closer to other cases than it is to other controls.  Distributions
of $S$ for cases (red) and controls (black) are given to the
right.  A significant shift toward higher $S$ values is seen in the
cases.  Odds ratios and FDR-adjusted OR $p$ values
are given.} 
}

\newcommand{\topcap}{
{Union of top three pathways.}
{SNPs from the top three pathways are combined to compute $S$ for the breast cancer data (a) and the liver cancer data (b).  Distributions
of $S$ for cases (red) and controls (black) are given to the
right.  A significant shift toward higher $S$ values is seen in the
cases.}
}


\ifthenelse{\boolean{withfigs}}
{
}{
	\clearpage 
	\section*{Figure Captions}
}

\flexfig{width=2.5in}{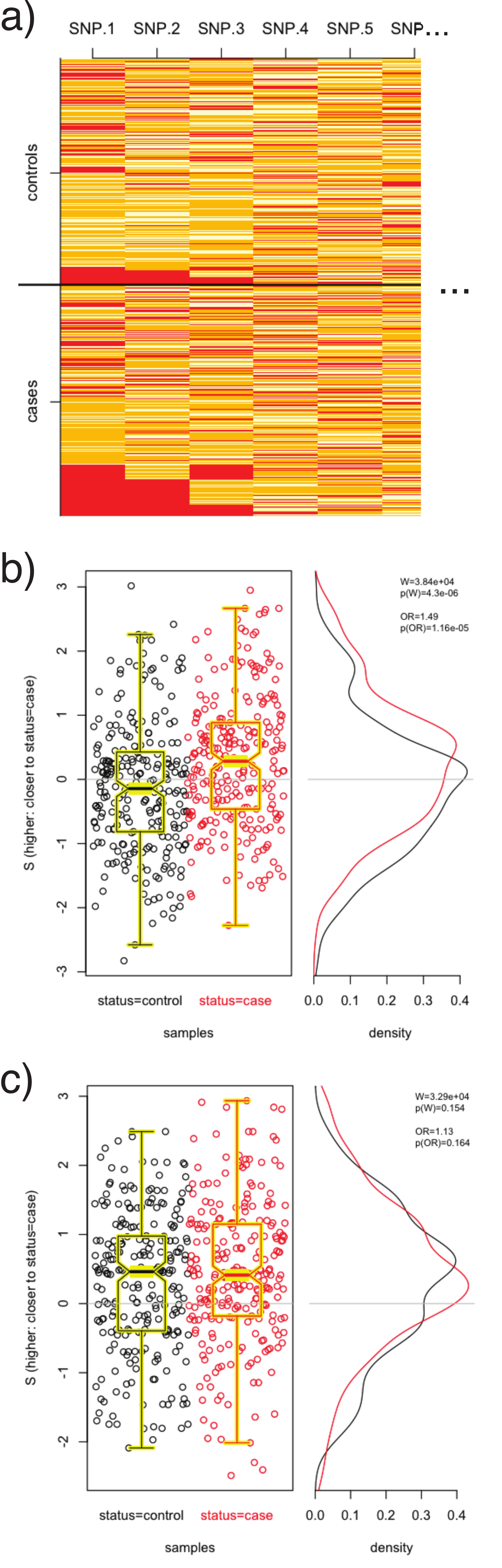}{exres}{\exrescap}
\flexfig{width=4in}{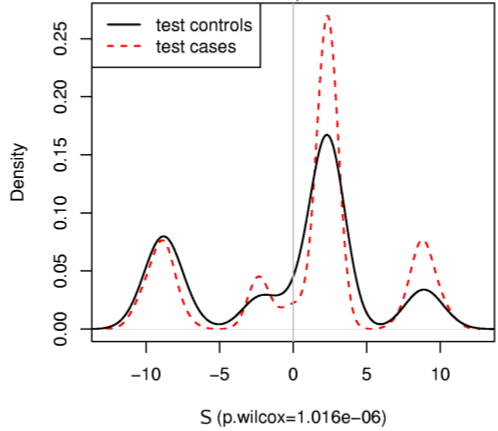}{multisnp-4fgfr2}{\fgfrcap}
\flexfig{width=6in}{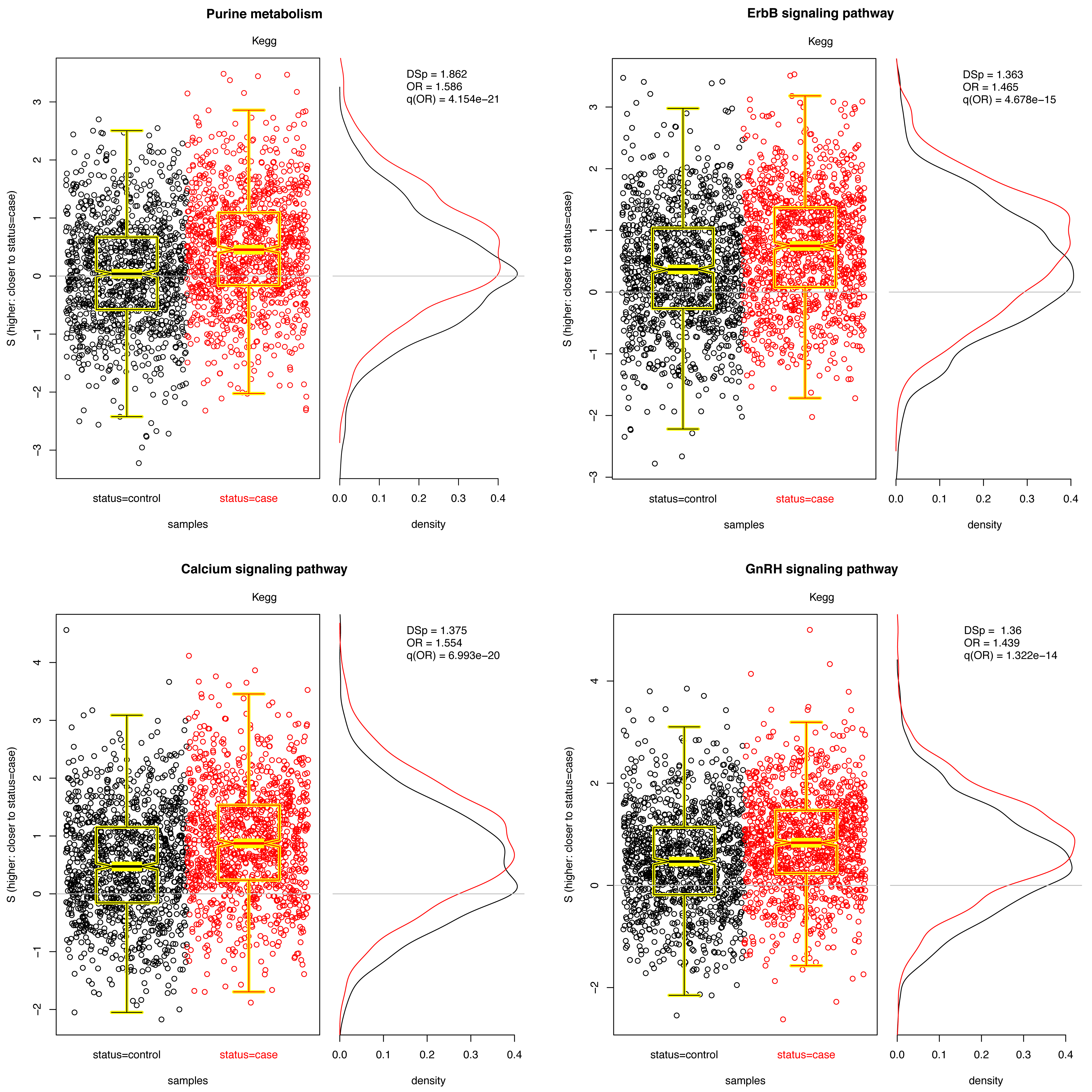}{brall}{\brallcap}
\flexfig{width=6in}{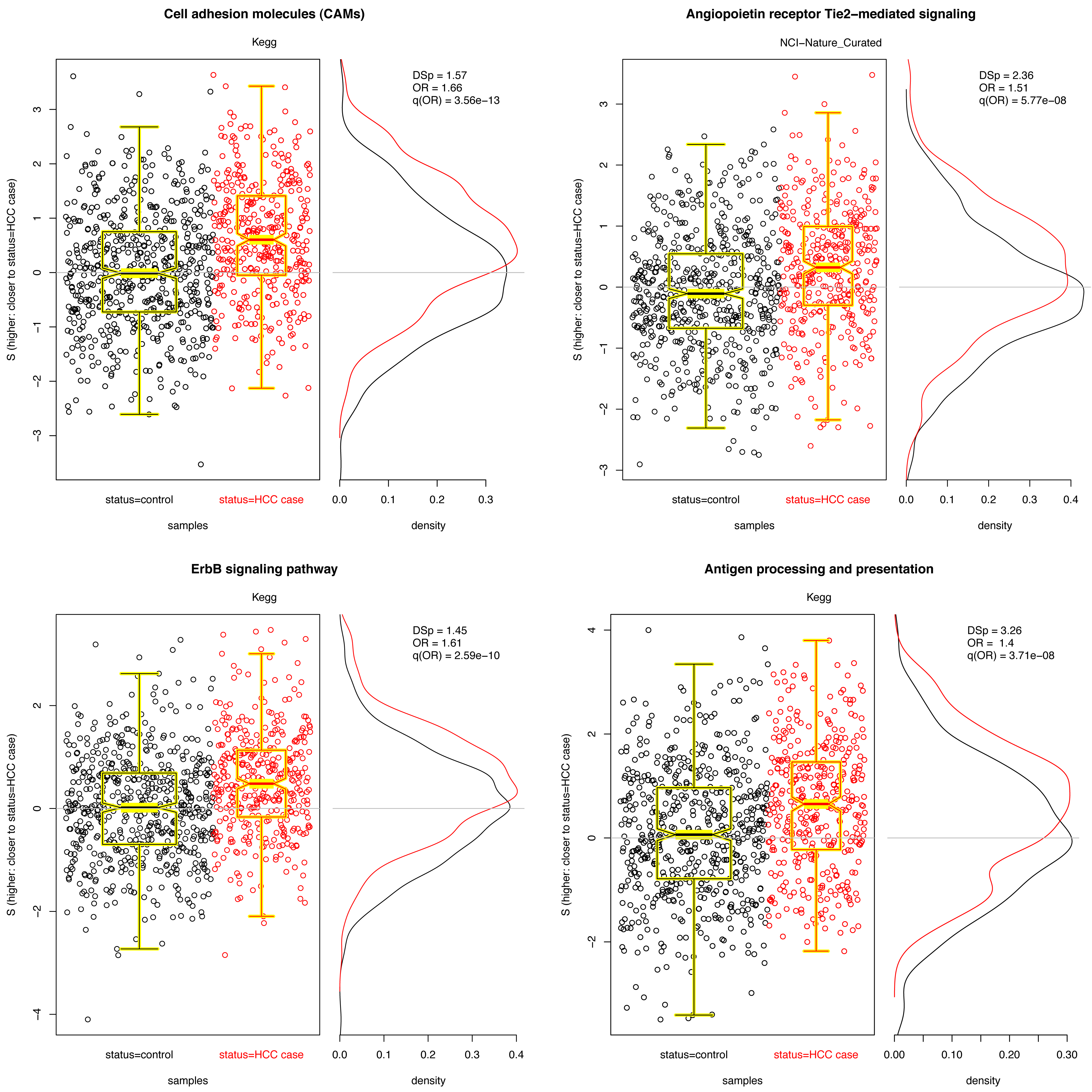}{livall}{\livallcap}
\flexfig{width=3in}{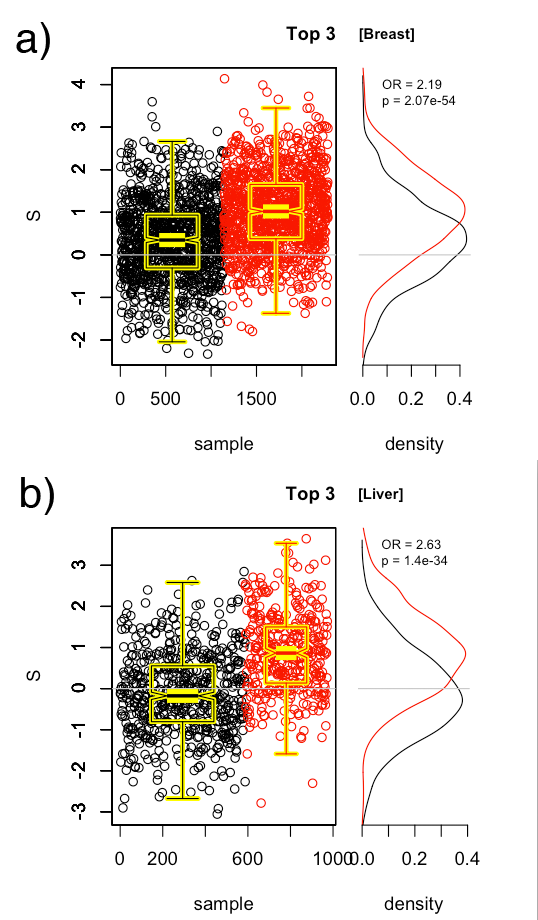}{top}{\topcap}

\clearpage
\begin{table}
\begin{center}
\begin{tabular}{ll}
\hline
&\textbf{Procedure for Pathways of Distinction Analysis}\\
\hline
1.&For a each pathway $P$, select all associated genes from pathway database such as PID~\cite{PID};\\
2.&For each gene on the pathway, select associated SNPs (e.g., using dbSNP) and choose the one\\
  &with the strongest association with case status, determined using Fisher's exact test;\\
3.&For each sample $Y$ in the GWAS, select the controls $F$ and cases $G$ which do not include it, \\
  &compute MAFs $f_i$ and $g_i$ for the SNPs $i$ selected in step 2, and compute $S_{Y,P}$ for each sample $Y$;\\
4.&Compare the distribution of $S_{Y,P}$ obtained in step 2 for cases to that of controls by computing\\
	&the Wilcoxon statistic $W_P$ based on the $S_{Y,P}$ for that pathway;\\
5.&Repeat steps 2--5 using permuted case/control labels, and normalize $W_P$ by the distribution\\
& of $W^*_P$ obtained with permuted labels, yielding the distinction score $DS_P$;\\
6.&Compare the distinction score $DS_P$ obtained in step 5 to that obtained for random sets of $l_P$ genes,\\ 
  &where $l_P$ is the number of genes in the pathway of interest.\\
\hline
\end{tabular}
\end{center}
\caption{ \label{alg1}Procedure for Pathways of Distinction Analysis}
\end{table}

\clearpage
\begin{table}[ht]
\begin{center}
{\tiny{\begin{tabular}{llrrrrr}
  \hline
Pathway & Source & Length & $DS_P$ & $p(DS_P)$ & O.R. & $q$(O.R.)\\
  \hline
Purine metabolism & Kegg & 136 & 1.86 & 6.36e-03 & 1.59 & 4.15e-21 \\ 
	Calcium signaling pathway & Kegg & 100 & 1.38 & 1.82e-03 & 1.55 & 6.99e-20 \\ 
  Melanogenesis & Kegg &  84 & 2.36 & 4.55e-03 & 1.53 & 1.47e-18 \\ 
  Gap junction & Kegg &  80 & 1.54 & 5.45e-03 & 1.49 & 1.49e-16 \\ 
  ErbB signaling pathway & Kegg &  81 & 1.36 & 1.45e-02 & 1.46 & 4.68e-15 \\ 
  Long-term potentiation & Kegg &  60 & 1.71 & 9.09e-04 & 1.45 & 4.34e-15 \\ 
  GnRH signaling pathway & Kegg &  79 & 1.36 & 1.18e-02 & 1.44 & 1.32e-14 \\ 
  TCR signaling in naive CD4+ T cells & NCI-Nature &  60 & 2.11 & 5.45e-03 & 1.42 & 7.80e-13 \\ 
  Prostate cancer & Kegg &  75 & 1.45 & 4.09e-02 & 1.38 & 4.37e-11 \\ 
  PKC-catalyzed phosphorylation \dots myosin phosphatase & BioCarta &  20 & 1.97 & $<$1e-04 & 1.30 & 5.82e-09 \\ 
  CCR3 signaling in eosinophils & BioCarta &  21 & 1.59 & 1.09e-02 & 1.29 & 8.86e-08 \\ 
  Biosynthesis of unsaturated fatty acids & Kegg &  18 & 1.69 & 2.45e-02 & 1.26 & 1.38e-06 \\ 
  Attenuation of GPCR signaling & BioCarta &  11 & 1.75 & 1.09e-02 & 1.25 & 2.41e-06 \\ 
  Stathmin and breast cancer resistance to antimicrotubule agents & BioCarta &  18 & 1.84 & 4.82e-02 & 1.24 & 4.96e-06 \\ 
  Visual signal transduction: Cones & NCI-Nature &  20 & 1.56 & 4.73e-02 & 1.24 & 2.24e-06 \\ 
  Dentatorubropallidoluysian atrophy (DRPLA) & Kegg &  11 & 1.84 & 2.73e-03 & 1.24 & 2.24e-06 \\ 
  Intrinsic prothrombin activation pathway & BioCarta &  22 & 1.35 & 3.18e-02 & 1.23 & 4.61e-06 \\ 
  Eicosanoid metabolism & BioCarta &  19 & 1.69 & 1.91e-02 & 1.23 & 3.44e-06 \\ 
  Effects of botulinum toxin & NCI-Nature &   7 & 1.44 & 2.27e-02 & 1.20 & 3.50e-05 \\ 
  Activation of PKC through G-protein coupled receptors & BioCarta &  10 & 1.50 & 9.09e-03 & 1.20 & 8.42e-06 \\ 
  Streptomycin biosynthesis & Kegg &   9 & 1.36 & 3.55e-02 & 1.17 & 1.89e-04 \\ 
  PECAM1 interactions & Reactome &   6 & 2.70 & 5.45e-03 & 1.17 & 7.28e-05 \\ 
  HDL-mediated lipid transport & Reactome &   8 & 1.47 & 2.00e-02 & 1.14 & 1.59e-03 \\ 
  Granzyme A mediated apoptosis pathway & BioCarta &   8 & 1.97 & 1.73e-02 & 1.12 & 6.60e-04 \\ 
\hline
\end{tabular}}}
\end{center}
\caption{\label{brtab}
PID pathways with significant $DS_P$
in the CGEMS breast cancer GWAS. (Pathways with over 60\% SNPs covered by another pathway have been removed; for the complete list, see Supplemental Table~\ref{brsup}). Pathway-length based resampled $p$-values, denoted $p(DS_P)$, are given for significant pathways, along with the odds ratios and associated FDRs for a logistic regression model.}
\end{table}

\clearpage
\begin{table}[ht]
\begin{center}
{\tiny{\begin{tabular}{llrrrrr}
  \hline
Pathway & Source & Length & $DS_P$ & $p(DS_P)$ & O.R. & $q$(O.R.)\\
  \hline
	Cell adhesion molecules (CAMs) & Kegg &  86 & 1.57 & 9.09e-03 & 1.66 & 3.56e-13 \\ 
  ErbB signaling pathway & Kegg &  76 & 1.45 & 3.45e-02 & 1.61 & 2.59e-10 \\ 
  Signaling events mediated by Stem cell factor receptor (c-Kit) & NCI-Nature &  40 & 2.35 & 5.45e-03 & 1.58 & 7.31e-10 \\ 
  Neurotrophic factor-mediated Trk receptor signaling & NCI-Nature &  50 & 1.60 & 2.36e-02 & 1.55 & 2.49e-08 \\ 
  Lissencephaly gene (LIS1) in neuronal migration and development & NCI-Nature &  21 & 2.02 & 7.27e-03 & 1.52 & 1.44e-07 \\ 
  Angiopoietin receptor Tie2-mediated signaling & NCI-Nature &  40 & 2.36 & 1.36e-02 & 1.51 & 5.77e-08 \\ 
  Reelin signaling pathway & NCI-Nature &  28 & 1.62 & 5.45e-03 & 1.46 & 7.35e-08 \\ 
  Syndecan-4-mediated signaling events & NCI-Nature &  27 & 1.74 & 1.64e-02 & 1.46 & 1.19e-06 \\ 
  Galactose metabolism & Kegg &  19 & 1.65 & 2.27e-02 & 1.44 & 5.01e-06 \\ 
  Vibrio cholerae infection & Kegg &  35 & 1.84 & 2.64e-02 & 1.43 & 6.67e-07 \\ 
  Paxillin-independent events mediated by a4b1 and a4b7 & NCI-Nature &  19 & 2.14 & 1.00e-02 & 1.40 & 6.67e-07 \\ 
  Antigen processing and presentation & Kegg &  34 & 3.26 & 1.36e-02 & 1.40 & 3.71e-08 \\ 
  Corticosteroids and Cardioprotection & BioCarta &  21 & 1.98 & 3.55e-02 & 1.39 & 1.24e-05 \\ 
  Lissencephaly gene (Lis1) in neuronal migration and development & BioCarta &  15 & 1.60 & 1.36e-02 & 1.37 & 2.52e-05 \\ 
  IL12 signaling mediated by STAT4 & NCI-Nature &  25 & 1.93 & 4.55e-02 & 1.37 & 1.58e-05 \\ 
  Biosynthesis of unsaturated fatty acids & Kegg &  13 & 1.76 & 1.64e-02 & 1.36 & 6.44e-05 \\ 
  Growth hormone signaling pathway & BioCarta &  18 & 1.75 & 3.18e-02 & 1.36 & 7.46e-05 \\ 
  Canonical Wnt signaling pathway & NCI-Nature &  28 & 1.92 & 4.73e-02 & 1.35 & 9.36e-06 \\ 
  NO2-dependent IL-12 pathway in NK cells & BioCarta &   8 & 1.82 & 2.73e-03 & 1.32 & 5.83e-05 \\ 
  Signaling events mediated by HDAC Class III & NCI-Nature &  19 & 2.12 & 3.91e-02 & 1.32 & 4.19e-05 \\ 
  Removal of aminoterminal propeptides from $\gamma$-carboxylated proteins & Reactome &   7 & 3.12 & 5.45e-03 & 1.29 & 8.46e-05 \\ 
  Aminophosphonate metabolism & Kegg &  13 & 1.91 & 3.36e-02 & 1.26 & 8.17e-04 \\ 
  Antigen processing and presentation & BioCarta &   6 & 2.61 & 1.82e-03 & 1.22 & 3.36e-05 \\ 
  Classical complement pathway & BioCarta &  12 & 2.27 & 1.55e-02 & 1.19 & 1.67e-04 \\ 
  Chylomicron-mediated lipid transport & Reactome &   7 & 1.94 & 3.27e-02 & 1.16 & 1.49e-02 \\ 
\hline
\end{tabular}}}
\end{center}
\caption{\label{livtab}
PID pathways with significant $DS_P$
in the liver cancer GWAS. (Pathways with over 60\% SNPs covered by another pathway have been removed; for the complete list, see Supplemental Table~\ref{livsup}). Pathway-length based resampled $p$-values, denoted $p(DS_P)$, are given for significant pathways, along with the odds ratios and associated FDRs for a logistic regression model.}
\end{table}

\clearpage
\begin{table}[ht]
\begin{center}
{\tiny{\begin{tabular}{lrrrr}
  \hline
Pathway & Length & $p(DS_P)$ & O.R. & $q$(O.R.) \\
  \hline
 Top-2 & 318 & $<$1e-04 & 2.02 & 1.63e-46 \\ 
 Top-3 & 397 & 1.00e-04 & 2.19 & 2.07e-54 \\ 
 Top-4 & 474 & $<$1e-04 & 2.33 & 3.65e-62 \\ 
 Top-5 & 522 & $<$1e-04 & 2.45 & 6.83e-66 \\ 
 Top-6 & 544 & $<$1e-04 & 2.44 & 8.51e-66 \\ 
 Top-7 & 558 & 2.00e-04 & 2.47 & 1.22e-67 \\ 
 Top-8 & 626 & $<$1e-04 & 2.59 & 1.01e-73 \\ 
 Top-9 & 658 & $<$1e-04 & 2.64 & 9.84e-75 \\ 
 Top-10 & 700 & $<$1e-04 & 2.77 & 9.72e-79 \\ 
 Top-11 & 710 & $<$1e-04 & 2.80 & 1.42e-79 \\ 
 Top-12 & 723 & $<$1e-04 & 2.82 & 2.06e-80 \\ 
 Top-13 & 739 & $<$1e-04 & 2.89 & 3.31e-82 \\ 
 Top-14 & 744 & $<$1e-04 & 2.93 & 2.86e-83 \\ 
 Top-15 & 770 & $<$1e-04 & 2.96 & 6.41e-85 \\ 
 Top-16 & 774 & $<$1e-04 & 2.97 & 5.10e-85 \\ 
 Top-17 & 791 & $<$1e-04 & 2.95 & 2.43e-85 \\ 
 Top-18 & 800 & $<$1e-04 & 3.06 & 1.15e-87 \\ 
 Top-19 & 814 & $<$1e-04 & 3.14 & 1.19e-89 \\ 
 Top-20 & 832 & $<$1e-04 & 3.26 & 4.51e-92 \\ 
 Top-21 & 837 & $<$1e-04 & 3.28 & 2.92e-92 \\ 
 Top-22 & 839 & $<$1e-04 & 3.29 & 2.41e-92 \\ 
 Top-23 & 845 & $<$1e-04 & 3.34 & 1.45e-93 \\ 
 Top-24 & 854 & $<$1e-04 & 3.38 & 4.62e-95 \\ 
\hline
\end{tabular}}}
\end{center}
\caption{\label{brtop}
PoDA results for sucessive unions of significant pathways in the CGEMS breast cancer data. Pathway-length based resampled $p$ values, denoted$p(DS_P)$, are given along with the odds ratios and associated FDRs for a logistic regression model.}
\end{table}

\clearpage
\begin{table}[ht]
\begin{center}
{\tiny{\begin{tabular}{lrrrr}
  \hline
Pathway & Length & $p(DS_P)$ & O.R. & $q$(O.R.) \\
  \hline
 Top-2 & 321 & 5.38e-02 & 2.37 & 1.20e-27 \\ 
 Top-3 & 402 & 2.80e-03 & 2.63 & 1.40e-34 \\ 
 Top-4 & 474 & 1.10e-03 & 2.86 & 6.50e-38 \\ 
 Top-5 & 539 & 9.00e-04 & 3.22 & 4.03e-42 \\ 
 Top-6 & 560 & 1.00e-04 & 3.39 & 1.19e-43 \\ 
 Top-7 & 580 & $<$1e-04 & 3.50 & 1.39e-44 \\ 
 Top-8 & 589 & 6.00e-04 & 3.50 & 1.35e-44 \\ 
 Top-9 & 603 & 4.00e-04 & 3.52 & 1.23e-44 \\ 
 Top-10 & 624 & $<$1e-04 & 3.60 & 1.33e-45 \\ 
 Top-11 & 640 & $<$1e-04 & 3.73 & 3.69e-47 \\ 
 Top-12 & 646 & $<$1e-04 & 3.78 & 1.68e-47 \\ 
 Top-13 & 667 & $<$1e-04 & 3.81 & 9.29e-48 \\ 
 Top-14 & 709 & 3.00e-04 & 3.88 & 1.90e-48 \\ 
 Top-15 & 751 & $<$1e-04 & 4.09 & 2.11e-49 \\ 
 Top-16 & 761 & $<$1e-04 & 4.09 & 1.76e-49 \\ 
 Top-17 & 797 & $<$1e-04 & 4.45 & 1.29e-50 \\ 
 Top-18 & 805 & $<$1e-04 & 4.46 & 5.24e-51 \\ 
 Top-19 & 823 & $<$1e-04 & 4.56 & 2.20e-51 \\ 
 Top-20 & 838 & $<$1e-04 & 4.56 & 1.73e-51 \\ 
\hline
\end{tabular}}}
\end{center}
\caption{\label{livtop}
PoDA results for sucessive unions of significant pathways in the liver cancer data. Pathway-length based resampled $p$ values, denoted $p(DS_P)$, are given along with the odds ratios and associated FDRs for a logistic regression model.}
\end{table}

\setcounter{table}{0}
\renewcommand{\thetable}{S-\arabic{table}}
\clearpage
\begin{table}[ht]
\begin{center}
{\tiny{\begin{tabular}{llrrrrr}
  \hline
Pathway & Source & Length & $DS_P$ & $p(DS_P)$ & O.R. & $q$(O.R.)\\
  \hline
	Purine metabolism & Kegg & 136 & 1.86 & 6.36e-03 & 1.59 & 4.15e-21 \\ 
  Calcium signaling pathway & Kegg & 100 & 1.38 & 1.82e-03 & 1.55 & 6.99e-20 \\ 
  Melanogenesis & Kegg &  84 & 2.36 & 4.55e-03 & 1.53 & 1.47e-18 \\ 
  Gap junction & Kegg &  80 & 1.54 & 5.45e-03 & 1.49 & 1.49e-16 \\ 
  ErbB signaling pathway & Kegg &  81 & 1.36 & 1.45e-02 & 1.46 & 4.68e-15 \\ 
  Long-term potentiation & Kegg &  60 & 1.71 & 9.09e-04 & 1.45 & 4.34e-15 \\ 
  GnRH signaling pathway & Kegg &  79 & 1.36 & 1.18e-02 & 1.44 & 1.32e-14 \\ 
  TCR signaling in naive CD4+ T cells & NCI-Nature &  60 & 2.11 & 5.45e-03 & 1.42 & 7.80e-13 \\ 
  TCR signaling in naive CD8+ T cells & NCI-Nature &  48 & 2.03 & 7.27e-03 & 1.38 & 1.11e-11 \\ 
  Prostate cancer & Kegg &  75 & 1.45 & 4.09e-02 & 1.38 & 4.37e-11 \\ 
  PKC-catalyzed phosphorylation \dots myosin phosphatase & BioCarta &  20 & 1.97 & $<$1e-04 & 1.30 & 5.82e-09 \\ 
  CCR3 signaling in eosinophils & BioCarta &  21 & 1.59 & 1.09e-02 & 1.29 & 8.86e-08 \\ 
  Biosynthesis of unsaturated fatty acids & Kegg &  18 & 1.69 & 2.45e-02 & 1.26 & 1.38e-06 \\ 
  Attenuation of GPCR signaling & BioCarta &  11 & 1.75 & 1.09e-02 & 1.25 & 2.41e-06 \\ 
  Stathmin and breast cancer resistance to antimicrotubule agents & BioCarta &  18 & 1.84 & 4.82e-02 & 1.24 & 4.96e-06 \\ 
  Visual signal transduction: Cones & NCI-Nature &  20 & 1.56 & 4.73e-02 & 1.24 & 2.24e-06 \\ 
  Dentatorubropallidoluysian atrophy (DRPLA) & Kegg &  11 & 1.84 & 2.73e-03 & 1.24 & 2.24e-06 \\ 
  Intrinsic prothrombin activation pathway & BioCarta &  22 & 1.35 & 3.18e-02 & 1.23 & 4.61e-06 \\ 
  Eicosanoid metabolism & BioCarta &  19 & 1.69 & 1.91e-02 & 1.23 & 3.44e-06 \\ 
  Effects of botulinum toxin & NCI-Nature &   7 & 1.44 & 2.27e-02 & 1.20 & 3.50e-05 \\ 
  Activation of PKC through G-protein coupled receptors & BioCarta &  10 & 1.50 & 9.09e-03 & 1.20 & 8.42e-06 \\ 
  Ca-calmodulin-dependent protein kinase activation & BioCarta &   8 & 1.70 & 1.00e-02 & 1.19 & 5.67e-05 \\ 
  Streptomycin biosynthesis & Kegg &   9 & 1.36 & 3.55e-02 & 1.17 & 1.89e-04 \\ 
  PECAM1 interactions & Reactome &   6 & 2.70 & 5.45e-03 & 1.17 & 7.28e-05 \\ 
  HDL-mediated lipid transport & Reactome &   8 & 1.47 & 2.00e-02 & 1.14 & 1.59e-03 \\ 
  Granzyme A mediated apoptosis pathway & BioCarta &   8 & 1.97 & 1.73e-02 & 1.12 & 6.60e-04 \\ 
\hline
\end{tabular}}}
\end{center}
\caption{\label{brsup}
Full list PID pathways with significant $DS_P$
in the breast cancer GWAS, including highly ``overlapping'' pathways. Pathway-length based resampled $p$-values, denoted $p(DS_P)$, are given for significant pathways, along with the odds ratios and associated FDRs for a logistic regression model.}
\end{table}

\clearpage
\begin{table}[ht]
\begin{center}
{\tiny{\begin{tabular}{llrrrrr}
  \hline
Pathway & Source & Length & $DS_P$ & $p(DS_P)$ & O.R. & $q$(O.R.)\\
  \hline
	Cell adhesion molecules (CAMs) & Kegg &  86 & 1.57 & 9.09e-03 & 1.66 & 3.56e-13 \\ 
  ErbB signaling pathway & Kegg &  76 & 1.45 & 3.45e-02 & 1.61 & 2.59e-10 \\ 
  Signaling events mediated by Stem cell factor receptor (c-Kit) & NCI-Nature &  40 & 2.35 & 5.45e-03 & 1.58 & 7.31e-10 \\ 
  Neurotrophic factor-mediated Trk receptor signaling & NCI-Nature &  50 & 1.60 & 2.36e-02 & 1.55 & 2.49e-08 \\ 
  Lissencephaly gene (LIS1) in neuronal migration and development & NCI-Nature &  21 & 2.02 & 7.27e-03 & 1.52 & 1.44e-07 \\ 
  Angiopoietin receptor Tie2-mediated signaling & NCI-Nature &  40 & 2.36 & 1.36e-02 & 1.51 & 5.77e-08 \\ 
  Reelin signaling pathway & NCI-Nature &  28 & 1.62 & 5.45e-03 & 1.46 & 7.35e-08 \\ 
  Syndecan-4-mediated signaling events & NCI-Nature &  27 & 1.74 & 1.64e-02 & 1.46 & 1.19e-06 \\ 
  Galactose metabolism & Kegg &  19 & 1.65 & 2.27e-02 & 1.44 & 5.01e-06 \\ 
  TPO signaling pathway & BioCarta &  17 & 2.61 & 6.36e-03 & 1.44 & 3.80e-06 \\ 
  Vibrio cholerae infection & Kegg &  35 & 1.84 & 2.64e-02 & 1.43 & 6.67e-07 \\ 
  Paxillin-independent events mediated by a4b1 and a4b7 & NCI-Nature &  19 & 2.14 & 1.00e-02 & 1.40 & 6.67e-07 \\ 
  Antigen processing and presentation & Kegg &  34 & 3.26 & 1.36e-02 & 1.40 & 3.71e-08 \\ 
  Corticosteroids and cardioprotection & BioCarta &  21 & 1.98 & 3.55e-02 & 1.39 & 1.24e-05 \\ 
  Lissencephaly gene (Lis1) in neuronal migration and development & BioCarta &  15 & 1.60 & 1.36e-02 & 1.37 & 2.52e-05 \\ 
  IL12 signaling mediated by STAT4 & NCI-Nature &  25 & 1.93 & 4.55e-02 & 1.37 & 1.58e-05 \\ 
  Biosynthesis of unsaturated fatty acids & Kegg &  13 & 1.76 & 1.64e-02 & 1.36 & 6.44e-05 \\ 
  Growth hormone signaling pathway & BioCarta &  18 & 1.75 & 3.18e-02 & 1.36 & 7.46e-05 \\ 
  Canonical Wnt signaling pathway & NCI-Nature &  28 & 1.92 & 4.73e-02 & 1.35 & 9.36e-06 \\ 
  NO2-dependent IL-12 pathway in nk cells & BioCarta &   8 & 1.82 & 2.73e-03 & 1.32 & 5.83e-05 \\ 
  Signaling events mediated by HDAC Class III & NCI-Nature &  19 & 2.12 & 3.91e-02 & 1.32 & 4.19e-05 \\ 
  Removal of aminoterminal propeptides from gamma-carboxylated proteins & Reactome &   7 & 3.12 & 5.45e-03 & 1.29 & 8.46e-05 \\ 
  Gamma-carboxylation, transport, and amino-terminal cleavage of proteins & Reactome &   6 & 3.25 & 1.82e-03 & 1.28 & 6.64e-05 \\ 
  Transport of $\gamma$-carboxylated protein precursors \dots & Reactome &   6 & 3.25 & 1.82e-03 & 1.28 & 6.64e-05 \\ 
  Paxillin-dependent events mediated by a4b1 & NCI-Nature &  17 & 1.84 & 2.00e-02 & 1.28 & 3.41e-05 \\ 
  Gamma-carboxylation of protein precursors & Reactome &   7 & 2.86 & 3.64e-03 & 1.28 & 1.38e-04 \\ 
  Aminophosphonate metabolism & Kegg &  13 & 1.91 & 3.36e-02 & 1.26 & 8.17e-04 \\ 
  Antigen processing and presentation & BioCarta &   6 & 2.61 & 1.82e-03 & 1.22 & 3.36e-05 \\ 
  Lectin induced complement pathway & BioCarta &  11 & 1.91 & 2.18e-02 & 1.20 & 1.55e-04 \\ 
  Classical complement pathway & BioCarta &  12 & 2.27 & 1.55e-02 & 1.19 & 1.67e-04 \\ 
  Chylomicron-mediated lipid transport & Reactome &   7 & 1.94 & 3.27e-02 & 1.16 & 1.49e-02 \\ 
\hline
\end{tabular}}}
\end{center}
\caption{\label{livsup}
Full list PID pathways with significant $DS_P$
in the liver cancer GWAS, including highly ``overlapping'' pathways. Pathway-length based resampled $p$-values, denoted $p(DS_P)$, are given for significant pathways, along with the odds ratios and associated FDRs for a logistic regression model.}
\end{table}

\end{document}